\documentclass[a4paper, twocolumn, fleqn]{article}
\usepackage{amsmath}
\numberwithin{equation}{section}
\title{\textbf{A charged space as the origin of sources, fields and potentials}}
\author{Koen J. van Vlaenderen \\
\emph{The Institute for Basic Research} \\
 email: nlx6461@nl.ibm.com}

\begin{document}

\maketitle

\begin{abstract}
The wave function $\psi$ is interpreted as charge density, or charge
distribution, at each point in space. This is a physical interpretation of $\psi$.
The notion of speed can be associated with $\psi$, which leads to the concept
of conduction currents and (displacement) convection currents.
The charge distribution is the origin of electrical and mechanical sources,
potentials and fields. The notion of \emph{self potential} is essential for defining
electrical or mechanical sources. Maxwell's equations are derived from the 
condition of charge conservation and mass conservation. There are two methods of 
modelling the mass of a charge: 
\begin{enumerate}
\item The mass of a charge is its electrostatic energy.
\item The mass of a charge is the energy of the Zero Point Field (ZPF) that interacts 
with the point charge.
\end{enumerate}
It is shown that the two models are related by a simple energy equation
for a particle at rest.
\end{abstract}


\section{Introduction}
An alternative study of microphysics \cite{Hofer}, called \emph{Material Wave Theory}
 (MTW), shows that the interpretation of $\psi$ as a physical wave is more realistic 
and more simple than the non-physical Copenhagen interpretation of $\psi$ 
representing only statistical qualities. The central conjecture of
MTW is the notion of \emph{intrinsic potential energy} of a particle.
This intrinsic potential energy turns out to be electromagnetic field energy.
Therefore, the wave nature of matter is closely related to electromagnetic energy.
There are several other theories that attribute even more importance to
electromagnetic phenomena.

\emph{Stochastic Electro-dynamics} (SED) \cite{SED} \cite{SED2} explains the statistical 
nature of micro-physics by a physical mechanism: quantum-like fluctuations of a random 
perturbing Zero Point Field. The inertia of a particle is described as a
reaction force that is a consequence of the anisotropy of the ZPF in an 
accelerating frame of reference. This means that inertia and also gravity are
secondary electromagnetic phenomena. The ZPF energy is described for the first
time as an extra term in the Planck (or blackbody) function.

In \cite{Petkov} the mass of a charge is considered equal to the electrostatic energy 
of the charge. An accelerating charge gives rise to a self force 
(Newton's reaction force), because the speed of light is anisotropic \cite{Petkov2}
around the accelerating charge. A local anisotropy of light speed is exactly 
a field of gravity. 

In \cite{Collins} the linear momentum of a charged particle is described as a 
self induced magnetic potential that acts on the charge, and also
rest-mass is considered here as the electrostatic energy.

Although these views have much in common (electrodynamics is essential in order to
explain the wave nature of matter, or to explain inertia and gravity),
it is not obvious how to integrate these theories into one consistent theory.
First, the electrodynamics of electrons within MWT is further explored.


\section{Classical Electrodynamics in MWT}
Hofer´s central conjecture within MWT \cite{Hofer} is: the intrinsic 
energy of a particle consists of kinetic \emph{and potential} energy.
Quantum Physics states that the intrinsic energy of a particle is solely kinetic.
This is not based on experiment, and the reasoning about the intrinsic nature 
of particles within the framework of Quantum Physics boils down to a logical circle
\cite{Hofer}~:

\emph{If a particle´s intrinsic energy is solely kinetic, the phase velocity of a de Broglie 
wave is not equal to the mechanical velocity of a particle. If phase velocity does not
equal mechanical velocity, a free particle can not exist of a single wave of specific
frequency and it must be formalized as a Fourier integral over infinitely many partial
waves. In this case any partial wave can not be interpreted as a physical wave.
Then the wave features of a partial wave can not be related to physical qualities. 
If they cannot be related to physical qualities, then internal processes must remain 
unconsidered. And if internal processes remain unconsidered, then the intrinsic energy 
of a particle is solely kinetic energy.}\\

Hofer´s conjecture is more simple and forces to describe the wave nature of
particles in terms of physical qualities, in stead of adopting the non-physical 
Hilbert space.


\subsection{The de Brogly wave}

First, the wave function $\psi$ is treated as a real function, in stead 
of a complex function, and with physical meaning:
\begin{equation}
\label{psi0}
\psi(\vec r,t) = \psi _0 \sin(\vec k \cdot \vec r - \omega t)
\end{equation}
\begin{equation}
\label{rho0}
\varrho(\vec r,t) = C\psi _0^2 \sin^2(\vec k \cdot \vec r - \omega t)
\end{equation}
The function $\varrho$ is the de Brogly wavefunction with dimension of
mass-density. A de Brogly wave is a \emph{mass oscillation}.
The periodic change of kinetic energy requires the existence of an
\emph{intrinsic} potential energy with a density of $\phi$.  The particle velocity
equals the phase velocity. By using an undefined constant C,
it is avoided to define the dimension of $\psi$. In this paper, C=1, 
and $\psi$ has the dimension of square root of mass-density.
These definitions define mechanical properties of $\psi$.

It is assumed that \emph{speed}, $\vec u$, can be associated 
with $\psi$ en $\varrho$~: $~~\vec p = \varrho \vec u~~$ is the impulse density and 
$~~w_{kin} = \frac{1}{2}\varrho u^2~~$ is the kinetic energy density $(u = |\vec u|)$.
A material wave is a periodic transformation of intrinsic kinetic energy 
and intrinsic potential energy, such that the sum of both intrinsic energy 
densities is constant:
\begin{equation}
\label{energy}
\frac{1}{2}\varrho u^2 + \phi = \phi_0 = constant
\end{equation}


\subsection{Electric and Magnetic Potentials}

In Material Wave theory it is shown that the intrinsic potential is electromagnetic
in nature. The definitions of the electric field and the magnetic
field, in terms of the intrinsic moment and intrinsic potential, are as
follows:
\begin{equation}
\vec E = -\nabla \frac{1}{\bar{\rho}}~\phi + \frac{1}{2\bar\rho}
\frac{\partial \vec p}{\partial t}
\label{EE}
\end{equation} 
\begin{equation}
\vec B = -\frac{1}{2\bar\rho}~\nabla \times \vec p
\label{BB}
\end{equation}
where $\bar \rho$ is a constant with the dimension of charge density
to guarantee compatibility with electromagnetic units. 
By substituting $\phi = \phi_0 - \frac{1}{2}\varrho u^2$ and $\vec p = \varrho \vec u$,
the equations \ref{EE} and \ref{BB} become:
\begin{equation}
\vec E = \nabla \frac{1}{2\bar\rho}~\varrho u^2 + \frac{1}{2\bar\rho}
\frac{\partial (\varrho \vec u )}{\partial t}
\label{E}
\end{equation} 
\begin{equation}
\vec B = -\frac{1}{2\bar\rho}~\nabla \times (\varrho \vec u)
\label{B}
\end{equation}
If an electric potential and magnetic potential are defined as follows:
\begin{equation}
\Phi = - \frac{\varrho}{2\bar\rho}~u^2 \qquad
\vec A = - \frac{\varrho}{2\bar\rho}~\vec u
\label{hofer}
\end{equation}
then it is obvious that the fields can be expressed in terms of the
potentials in the usual way. The sources $\rho _s$ and $\vec J_s$
can be expressed also in terms of the potentials $\Phi$ and $\vec A$.
\begin{equation}
\vec E = -\nabla \Phi - \frac{\partial \vec A}{\partial t}
\label{defE}
\end{equation}
\begin{equation}
\vec B = \nabla \times \vec A
\label{defB}
\end{equation}
\begin{equation}
\rho _s = \epsilon \left( \mu \epsilon \frac{\partial ^2 \Phi}{\partial t^2}
   - \nabla ^2 \Phi \right)
\end{equation}
\begin{equation}
\vec {J_s} = \frac{1}{\mu} \left( \mu \epsilon \frac{\partial ^2 \vec A}{\partial t^2}
   - \nabla ^2 \vec A \right)
\end{equation}


\subsection{Maxwell's equations}
\label{ME}
If mass is conserved and $\epsilon \mu = \frac{1}{u^2} = constant$ then
Maxwell's equations are valid in MWT. Proof: starting with the mass 
continuity equation, the Lorentz gauge can be derived:
\begin{equation}
\begin{split}
0 & =
\nabla \! \cdot \! (\varrho ~ \vec u) + \dfrac{\partial \varrho}{\partial t}\\
& = \nabla \! \cdot \! (- \dfrac{\varrho}{2\bar\rho}~ \vec u) +
  \mu\epsilon \dfrac{\partial \! \left( -\dfrac{\varrho}{2\bar\rho}~u^2 \right) }{\partial t}\\
& = \nabla \! \cdot \! \vec A + \mu\epsilon \dfrac{\partial \Phi}{\partial t}
\end{split}
\end{equation}
Maxwell's equations follow from the definitions of fields and sources and the
Lorentz gauge:

\begin{equation} \begin{split}
\nabla \times \vec E & = \nabla \times \left ( - \nabla \Phi - \dfrac{\partial \vec A}{\partial t} \right ) \\
& = - \dfrac{\partial \left ( \nabla \times \vec A \right )}{\partial t} 
  = - \dfrac{\partial \vec B}{\partial t}
\end{split} \end{equation}
\begin{equation} \begin{split}
\nabla \cdot \vec E & = \nabla \times \left ( - \nabla \Phi - \dfrac{\partial \vec A}{\partial t} \right ) \\
& = - \nabla^2 \Phi - \dfrac{\partial}{\partial t} \nabla \cdot \vec A \\
& = - \nabla^2 \Phi + \mu\epsilon \dfrac{\partial ^2 \Phi}{\partial t^2}
  = \dfrac{\rho _s}{\epsilon}
\end{split} \end{equation}
\begin{equation} \begin{split}
\nabla\times \vec B & = \nabla \times \nabla \times \vec A \\
& = \nabla (\nabla \cdot \vec A)- \nabla ^2 \vec A \\
& = - \nabla \left( \mu\epsilon \dfrac{\partial \Phi}{\partial t} \right) - \nabla ^2 \vec A \\
& = \mu\epsilon \dfrac{\partial}{\partial t}\left( \vec E + \dfrac{\partial \vec A}{\partial t} \right) - \nabla ^2 \vec A \\
& = \mu\epsilon \dfrac{\partial \vec E}{\partial t} + \left( \mu \epsilon \frac{\partial ^2 \vec A}{\partial t^2}  - \nabla ^2 \vec A \right) \\
& = \mu\epsilon \dfrac{\partial \vec E}{\partial t} + \mu \vec {J_s}
\end{split} \end{equation}
\begin{equation} \begin{split}
\nabla\cdot \vec B = \nabla \cdot \nabla \times \vec A = 0  
\end{split} \end{equation}
If $\vec u = constant$ then $\vec E \perp \vec u,
\quad \vec B \perp \vec u$ and $\vec E \perp \vec B$.
Proof: let $a = - \dfrac{\varrho}{2\bar\rho}$ and let $\vec g = \nabla a$;
$a$ is conserved, because $\varrho$ is conserved and $\bar \rho$ is a constant.
\begin{equation}\begin{split}
\vec B & = \nabla \times \vec A = \nabla \times (a \vec u) = (\nabla a)\times \vec u \\
& = \vec g \times \vec u
\end{split}\end{equation}
\begin{equation}\begin{split}
\vec E & = -\nabla \Phi - \dfrac{\partial \vec A}{\partial t} 
  = -u^2 \nabla a - \dfrac{\partial a}{\partial t}\vec u \\
& = -u^2 \vec g + \nabla \cdot(a \vec u)\vec u \\
& = -(\vec u \cdot \vec u)\vec g + (\vec g \cdot \vec u)\vec u \\
& = (\vec g \times \vec u)\times \vec u = \vec B \times \vec u  
\end{split}\end{equation}


\subsection{Field energy and Pointing vector}
\label{FP}
In case of $\vec u = constant$ the expressions of field energy
and pointing flow, in terms of $\vec g$ and $\vec u$, become:
\begin{equation}\begin{split}
\dfrac{\mu}{2}H^2 & = \dfrac{\mu}{2} \left( \dfrac{\vec B}{\mu} \cdot \dfrac{\vec B}{\mu} \right) \\
 & = \dfrac{1}{2 \mu}(\vec g \times \vec u) \cdot (\vec g \times \vec u) \\
 & = \dfrac{1}{2 \mu}|\vec g \times \vec u|^2
\end{split}\end{equation} 
\begin{equation}\begin{split}
\dfrac{\epsilon}{2} E^2 & = \dfrac{\epsilon}{2} (\vec B \times \vec u) \cdot (\vec B \times \vec u) \\
 & = \dfrac{\epsilon}{2}u^2(\vec B \cdot \vec B) = \dfrac{1}{2 \mu}|\vec g \times \vec u|^2 
\end{split}\end{equation}
\begin{equation}\begin{split}
\vec E \times \vec H & = \frac{1}{\mu}(\vec E \times \vec B) = \frac{1}{\mu}(\vec B \times \vec u)\times \vec B \\
 & = \frac{1}{\mu}B^2 \vec u - \frac{1}{\mu} (\vec B \cdot \vec u) \times \vec B \\
 & = \frac{1}{\mu}|\vec g \times \vec u|^2 \vec u
  = \left( \frac{\mu}{2}H^2 + \frac{\epsilon}{2}E^2 \right) \vec u
 \label{Pointing}
\end{split}\end{equation}
Equation \ref{Pointing} is Pointing's Theorem in Material Wave Theory. Notice that 
if $\vec g \times \vec u = 0$ then the field energies are zero, and also the energy flow 
is zero. The mass gradient of the matter wave must have a non-zero component that is
perpendicular to the direction of motion. Otherwise there is no intrinsic potential
energy. Therefore the monochromatic \emph{plane} particle wave (see equation 
\ref{rho0}) cannot be an adequate description of a matter wave with intrinsic kinetic 
energy and intrinsic electro-magnetic (potential) energy, because in this case
$\vec g \times \vec u = \vec 0$~!


\section{Self induced potentials}
At this point it is worthwhile to make a comparison with the notion
of the \emph{self-induced} magnetic potential of a charge \cite{Collins}:
\begin{equation}
m \vec u = q \vec A \quad \Rightarrow \quad \vec A = \dfrac{m}{q}~\vec u
~= \frac{\varrho}{\rho}~\vec u
\label{Collins}
\end{equation}
This equation is the result of combining Newton's laws with Maxwell's
equations, as follows: an applied force causes an elementary particle, 
with mass $m$ and charge $q$, to accelerate.
\begin{equation}
\vec F = m \vec a = \dfrac{\partial (m \vec v)}{\partial t}
\end{equation}
The term $-\frac{\partial \vec A}{\partial t}$ in equation \ref{defE}
is caused by the applied force. If the particle is not accelerated then
this term is zero. Therefore, $-q \frac{\partial \vec A}{\partial t}$,
which is an extra Coulomb force, is also Newton's reaction force $F'$.
\begin{equation}\begin{split}
F & = -F' \quad \Rightarrow \; \dfrac{\partial (m \vec v)}{\partial t} =
\dfrac{\partial (q \vec A) }{\partial t} \quad \Rightarrow \\
m \vec v & = q \vec A
\end{split}\end{equation}
A similar equation exists for the electric potential:
\begin{equation}
m c^2 = q \Phi \quad \Rightarrow \quad \Phi = \dfrac{m}{q}~c^2 = \dfrac{\varrho}{\rho}~c^2
\end{equation}
meaning the total energy of a charge is \emph{electrostatic}.
In equation \ref{hofer}, $\bar \rho$ is a constant, which is not the
case in equation \ref{Collins}, where $\rho$ is a \emph{scalar function}.
It is not clear why $\bar\rho$ is defined a constant, except for
compatibility between units for mechanical quantities and variables for
electromagnetical quantities. The simplest view is to consider the self-induced 
potentials and the intrinsic potentials of MWT as equal, and to be called
\emph{self potentials}. This means that we have to replace the constant 
$\bar \rho$ for the scalar $-\frac{1}{2}\rho$.


\subsection{The electromagnetic self potentials}
Since $\psi$ has real physical interpretation, the following question
comes to mind: is $\rho$ a function of $\psi$? If, for instance, $\rho = \psi$,
then space is filled with "charge", or even consists of "charge".
This model is in agreement with notions like \emph{displacements currents} or
\emph{convection displacement currents} \cite{Chubykalo}. Such a current has 
to exist beside conduction currents in order to solve a paradox in the 
Faraday-Maxwell theory. The definitions of the electric and magnetic potentials
(in case $-\frac{1}{2}\bar\rho = \rho = \psi$) becomes:
\begin{equation}
\Phi = \dfrac{\varrho}{\rho}u^2 = \dfrac{\psi^2}{\psi}u^2 = \psi u^2
\end{equation}
\begin{equation}
\vec A = \dfrac{\varrho}{\rho}\vec u = \dfrac{\psi^2}{\psi}\vec u = \psi \vec u
\end{equation}
The definitions of electromagnetic sources and fields now become:
\begin{equation} 
\vec E = -\nabla \psi u^2 - \dfrac{\partial(\psi \vec u)}{\partial t}
\end{equation}
\begin{equation}
\vec B = \nabla \times (\psi \vec u)
\end{equation}
\begin{equation}
\rho _s = \epsilon \left( \mu \epsilon \frac{\partial ^2(\psi u^2)}{\partial t^2}
   - \nabla ^2(\psi u^2) \right)
\label{chargesource}
\end{equation}
\begin{equation}
\vec {J_s} = \frac{1}{\mu} \left( \mu \epsilon \frac{\partial ^2(\psi\vec u)}{\partial t^2}
   - \nabla ^2(\psi\vec u) \right)
\label{currentsource}
\end{equation}
It would be unnatural to distinguish between $\rho_s$ and $\rho (=\psi)$:
\begin{equation}
\psi = \epsilon \left( \mu \epsilon \frac{\partial ^2(\psi u^2)}{\partial t^2}
   - \nabla ^2(\psi u^2) \right)
\label{selfpotential1}
\end{equation}
\begin{equation}
\psi \vec u = \frac{1}{\mu} \left( \mu \epsilon \frac{\partial ^2(\psi\vec u)}{\partial t^2}
   - \nabla ^2(\psi\vec u) \right)
\label{selfpotential2}
\end{equation}
If $\psi$ satisfies equations \ref{selfpotential1} and \ref{selfpotential2} at some point
in space, then $\psi$ is a \emph{source} at that particular point.
Equations \ref{selfpotential1} and \ref{selfpotential2} are called the
\emph{self-potential equations}. They can be reformulated in terms of the potentials:
\begin{equation}
\Phi = \frac{1}{\mu} \left( \mu \epsilon \frac{\partial ^2\Phi}{\partial t^2}
   - \nabla ^2\Phi \right)
\label{selfpot1}
\end{equation}
\begin{equation}
\vec A = \frac{1}{\mu} \left( \mu \epsilon \frac{\partial ^2\vec A}{\partial t^2}
   - \nabla ^2\vec A \right)
\label{selfpot2}
\end{equation}
If charge is conserved and $\mu\epsilon =\frac{1}{u^2} = constant$,
then Maxwell's equations are valid.
The proof is similar to the proof in section \ref{ME}. First, the 
Lorentz gauge is derived from the charge continuity equation:
\begin{equation}    
0 = \nabla \! \cdot \! (\psi ~ \vec u) + \dfrac{\partial \psi}{\partial t}
\end{equation}
(Etc...). Substitute $\vec g = \nabla \psi$ and take $\vec u = constant$, then
the EM fields are perpendicular to each other (see section \ref{ME}).
Also the same expressions for the energy densities and Pointing vector 
can be derived by substituting $\vec g = \nabla \psi$ (see section \ref{FP}).  

Equation \ref{energy} ($\frac{1}{2}\varrho u^2 + \phi = \phi_0 = constant$),
can be rewritten in terms of electric energy density, magnetic energy density
and static electric energy density (which is the total energy):

\begin{equation}
\frac{1}{2}\varrho u^2 + \frac{\epsilon}{2}E^2 + \frac{\mu}{2}H^2 = \rho\Phi
\label{eeq}
\end{equation}


\subsection{The mechanical self potentials}
In analogy with electromagnetic sources, fields and potentials, one can define
mechanical sources, fields and potentials, simply by substituting $\psi^2$ for
$\psi$:
\begin{equation}
w = \psi^2 u^2 \quad \quad \vec p = \psi^2 \vec u
\end{equation}
\begin{equation} 
\vec f = \nabla w + \dfrac{\partial \vec p}{\partial t}
\end{equation}
\begin{equation}
\vec s = \nabla \times \vec p
\end{equation}
\begin{equation}
\varrho _s = \epsilon _m \left( \mu _m \epsilon _m \frac{\partial ^2 w}{\partial t^2}
   - \nabla ^2 w \right)
\end{equation}
\begin{equation}
\vec {p_s} = \frac{1}{\mu _m} \left( \mu _m \epsilon _m \frac{\partial ^2 \vec p}{\partial t^2}
   - \nabla ^2 \vec p \right)
\end{equation}
$w$ and $\vec p$ are the potential energy density and potential momentum density. The
vector fields $\vec f$ and $\vec s$ are the \emph{force density} field and
\emph{angular momentum density} field. An intrinsic angular momentum is also called
\emph{spin}, and therefore the symbol $s$ is used. 
The constants $\epsilon _m$ and $\mu _m$ are the mechanical analogies of
$\epsilon$ and $\mu$. The force density is zero if energy-momentum is conserved.

In case $\epsilon _m \mu _m = \frac{1}{u^2} = constant$ then also
for the mechanical fields the Maxwell's equations apply:
\begin{equation}
\nabla \times \vec f = \dfrac{\partial \vec s}{\partial t}
\label{spin}
\end{equation} 
\begin{equation}
\label{gravity}
\nabla \cdot \vec f = - \dfrac{\varrho _s}{\epsilon _m}
\end{equation}
\begin{equation}
\nabla \times \vec s = - \epsilon _m \mu _m \dfrac{\partial \vec f}{\partial t} +
                       \mu _m \vec {p _s}
\label{ampere}
\end{equation}
\begin{equation}
\nabla \cdot \vec s = 0
\end{equation}
Equation \ref{spin} expresses that the spin increases or decreases in case
the rotation of force density is not zero. Equation \ref{gravity} is the law
of gravity in differential form. Equation \ref{ampere} is the mechanical
equivalent of Amp\`ere's law.

Next, it is unnatural to distinguish between $\varrho _s$ and $\varrho = \psi^2$
and therefore we can speak also of the mechanical \emph{self potentials}:
\begin{equation}
w = \frac{1}{\mu _m} \left( \mu _m \epsilon _m \frac{\partial ^2 w}{\partial t^2}
   - \nabla ^2 w \right)
\end{equation}
\begin{equation}
\vec p = \frac{1}{\mu _m} \left( \mu _m \epsilon _m \frac{\partial ^2 \vec p}{\partial t^2}
   - \nabla ^2 \vec p \right)
\end{equation}
Surprisingly, equation \ref{eeq} can be derived by using the definition of the mechanical
self potential:\\
($\epsilon _m = \epsilon, \quad \mu _m = \mu, \quad \epsilon\mu = \frac{1}{u^2}, \quad \vec u = constant$)
\begin{equation}
\begin{split}
& \frac{1}{2}\varrho _s u^2 = \frac{1}{2} w 
  = \frac{1}{2\mu} \left( \epsilon \mu \frac{\partial^2 w}{\partial t^2} -
     \nabla ^2 w \right) \\
& =  \frac{1}{2\mu} \left[  \epsilon \mu 2u^2 \left( \frac{\partial \psi}{\partial t} \right)^2 
      + \epsilon \mu 2 \psi \frac{\partial ^2 (\psi u^2)}{\partial t^2} \right] + \\
& \quad \quad \frac{1}{2\mu} \left[ -2u^2(\nabla \psi)^2 -2\psi \nabla ^2(\psi u^2) \right] \\
& =  \frac{1}{\mu} \left[ (\nabla \cdot (\psi \vec u))^2 - u^2(\nabla \psi)^2 \right] + \\ 
     & \quad \quad \frac{\psi}{\mu} \left[ \epsilon \mu \frac{\partial^2(\psi u^2)}{\partial t^2}
       - \nabla ^2(\psi u^2) \right] \\
& = \frac{1}{\mu}[(\nabla \psi \cdot \vec u)^2 - (\nabla \psi)^2 u^2] + \psi \Phi \\
& = -\frac{1}{\mu}|\nabla \psi \times \vec u|^2 + \psi \Phi \\
& = -\frac{\mu}{2}H^2 - \frac{\epsilon}{2}E^2 + \psi \Phi
\end{split}
\end{equation}
The self potentials are defined such that the energy density equation \ref{eeq} is 
fulfilled. If we substitute $\varrho = \psi ^2$ and $\Phi = \psi u^2$ then we get:
\begin{equation}
\frac{1}{2} \psi ^2 u^2 = \frac{\mu}{2}H^2 + \frac{\epsilon}{2}E^2
\end{equation}

		   
\section{Physical Units}
Since $\psi = \dfrac{q}{V}$ and $\psi ^2 = \dfrac{m}{V}$, the unit Amp\`ere is no 
longer free for definition:
\begin{equation}
\begin{split}
\left[ \frac{Coulomb}{m^3} \right] ^2 & = \left[ \frac{Kg}{m^3} \right] \quad \Rightarrow \\
[Coulomb] & = [\sqrt{Kg~m^3}]
\end{split}
\end{equation}
\begin{equation}
[Amp\grave ere] = \left[ \frac{\sqrt{Kg~m^3}}{s} \right]
\end{equation}
\begin{equation}
[\epsilon] = [s^2] \qquad [\mu] = [m^{-2}]
\end{equation}
Other units (expressed in mechanical base units) are:\\
$[Volt] = \left[ \dfrac{\sqrt{Kg~m}}{s^2} \right]$, $\quad [\Omega] = [m^{-1}s^{-1}]$\\
$[Farad] = [s^2m], \quad [Henry] = [m^{-1}]$.\\

Especially, the definition of Coulomb is interesting. It seems that the spatial 
dimension of a charge is $\frac{3}{2}$. This is a fractal dimension. One might interpret
a charge as a point-like particle (without mass) that follows a trajectory with a fractal
dimension of $\frac{3}{2}$, within a closed volume.


\section{Discussion}
Equations 2.18 to 2.22 can be derived also from the the weaker pre-condition of
$\nabla \times \vec u = \vec 0$ and $\nabla \cdot \vec u = 0$, in stead of
$\vec u = constant$. A fractal trajectory within a closed volume is an example
of \\
$\vec u \neq constant, \quad \nabla \times \vec u = \vec 0, \quad \nabla \cdot \vec u = 0$.
A trajectory with a fractal dimension is in agreement with Stochastic Electro
Dynamics, because SED assumes a massless parton to interact with the ZPF that has
a \emph{broad} spectrum. 

Suppose, the massless parton has a speed $|\vec u| = c$, then equation \ref{eeq} becomes:
\begin{equation}
\frac{1}{2}\varrho c^2 + \frac{\epsilon}{2}E^2 + \frac{\mu}{2}H^2 = \rho\Phi
\end{equation}
Its total energy density is $\varrho c^2 = \rho\Phi$. This can only be understood by the
notion of (intrinsic) \emph{self potentials}, as introduced in this paper. If the
closed and finite volume that confines the parton's trajectory is motionless, then one
speaks of rest-energy or rest-mass. This combines the different models, as described
in the introduction, such that it yields one theory.

The charged field $\psi$ does not show Coulomb interaction or gravity interaction
between every two points. In other words: not all points in space are \emph{sources}.
Only those points in space that satisfy the self potential equations
can show Coulomb interaction or gravity interaction.
Thus, the charged space $\psi$ is the origin of sources and (self) potentials.

\end{document}